\documentclass[prb,twocolumn,showpacs,superscriptaddress]{revtex4}%
\usepackage{amsfonts}
\usepackage{amsmath}
\usepackage{amssymb}
\usepackage{graphicx}%
\providecommand{\U}[1]{\protect\rule{.1in}{.1in}}

\begin{document}
\title{Structural, mechanical, thermodynamic, and electronic properties of thorium hydrides from first principles}
\author{Bao-Tian Wang}
\affiliation{Institute of Theoretical Physics and Department of
Physics, Shanxi University, Taiyuan 030006, People's Republic of
China} \affiliation{LCP, Institute of Applied Physics and
Computational Mathematics, Beijing 100088, People's Republic of
China}
\author{Ping Zhang}
\thanks{Author to whom correspondence should be
addressed. E-mail: zhang\_ping@iapcm.ac.cn} \affiliation{LCP,
Institute of Applied Physics and Computational Mathematics, Beijing
100088, People's Republic of China} \affiliation{Center for Applied
Physics and Technology, Peking University, Beijing 100871, People's
Republic of China}
\author{Hongzhou Song}
\affiliation{LCP, Institute of Applied Physics and Computational
Mathematics, Beijing 100088, People's Republic of China}
\author{Hongliang Shi} \affiliation{LCP, Institute
of Applied Physics and Computational Mathematics, Beijing 100088,
People's Republic of China} \affiliation{SKLSM, Institute of
Semiconductors, Chinese Academy of Sciences, People's Republic of
China}
\author{Dafang Li}
\affiliation{LCP, Institute of Applied Physics and Computational
Mathematics, Beijing 100088, People's Republic of China}
\author{Wei-Dong Li}
\affiliation{Institute of Theoretical Physics and Department of Physics, Shanxi University,
Taiyuan 030006, People's Republic of China}

\begin{abstract}
We perform first-principles calculations of the structural,
electronic, mechanical, and thermodynamic properties of thorium
hydrides (ThH$_{2}$ and Th$_{4}$H$_{15}$) based on the density
functional theory with generalized gradient approximation. The
equilibrium geometries, the total and partial densities of states,
charge density, elastic constants, elastic moduli, Poisson's ratio,
and phonon dispersion curves for these materials are systematically
investigated and analyzed in comparison with experiments and
previous calculations. These results show that our calculated
equilibrium structural parameters are well consistent with
experiments. The Th$-$H bonds in all thorium hydrides exhibit weak
covalent character, but the ionic properties for ThH$_{2}$ and
Th$_{4}$H$_{15}$ are different due to their different hydrogen
concentration. It is found that while in ThH$_{2}$ about 1.5
electrons transfer from each Th atom to H, in Th$_{4}$H$_{15}$ the
charge transfer from each Th atom is around 2.1 electrons. Our
calculated phonon spectrum for the stable body-centered tetragonal
phase of ThH$_{2}$ accords well with experiments. In addition we
show that ThH$_{2}$ in the fluorite phase is mechanically and
dynamically unstable.
\end{abstract}
\maketitle

\section{INTRODUCTION}

Thorium is one kind of important nuclear materials and together with
its compounds has been widely investigated both experimentally and
theoretically. Among thorium compounds, thorium hydrides (ThH$_{2}$
and Th$_{4}$H$_{15}$) are metallic solids and have potential use for
advanced nuclear fuels. In addition, Th$_{4}$H$_{15}$ has been
reported to have superconductivity with transition temperatures
\emph{T}$_{c}$ $\mathtt{\sim}$8 K \cite{Satterthwaite1} and has been
considered as promising candidates for hydrogen storage since its
large hydrogen-to-metal ratio \cite{Shein}. ThH$_{2}$ is not
superconducting above 1 K \cite{Satterthwaite2}.

Despite the abundant research on thorium hydrides, relatively little
is known regarding their chemical bonding, mechanical properties,
and phonon dispersion. Only the optical phonon density of states of
ThH$_{2}$ and Th$_{4}$H$_{15}$ was measured through inelastic
neutron scattering \cite{Dietrich} and the bulk modulus of ThH$_{2}$
was calculated by linear muffin-tin orbital (LMTO) method
\cite{Brooks}. Until now the elastic properties, which relate to
various fundamental solid-state properties such as interatomic
potentials, equation of state, phonon spectra, and thermodynamical
properties, are unknown for ThH$_{2}$ and Th$_{4}$H$_{15}$. In
addition, although the electronic properties as well as the chemical
bonding in thorium hydrides have been calculated recently by Shein
\emph{et al.} \cite{Shein} through the full-potential LAPW (FLAPW)
method, the study of the bonding nature of Th$-$H bond involving its
mixed ionic/covalent character is still lacking. These facts, as a
consequence, inhibit deep understanding of thorium hydrides.
Motivated by these observations, in this paper, we present a
first-principles study by calculating the structural, electronic,
mechanical, and thermodynamic properties of thorium hydrides. Also,
the stability of the metastable phase of ThH$_{2}$ (fluorite
structure with space group \emph{Fm$\bar{3}$m}) is analyzed and our
calculated results show that the fluorite-type ThH$_{2}$ is
mechanically unstable. We perform the Bader analysis
\cite{Bader,Tang} of thorium hydrides and find that about 1.5 (2.1)
electrons transfer from each Th atom to H for ThH$_{2}$
(Th$_{4}$H$_{15}$).

\section{computational methods}

Our total energy calculations are carried out by employing the
plane-wave basis pseudopotential method as implemented in Vienna
\textit{ab initio} simulation package (VASP) \cite{Kresse3}. The
exchange and correlation effects are described by the density
functional theory (DFT) within generalized gradient approximation
(GGA) \cite{GGA}. The projected augmented wave (PAW) method of
Bl\"{o}chl \cite{PAW} is implemented in VASP with the frozen-core
approximation. The thorium 6s$^{2}$7s$^{2}$6p$^{6}$6d$^{1}$5f$^{1}$
and the hydrogen 1s$^{1}$ electrons are treated as valence
electrons. Note that although the $5f$ states are empty in elemental
Th, this level turns to evolve into a hybridization with the
hydrogen orbitals both in the valence band and the conduction band,
as well as to prominently contribute to the conduction band (see
Fig. \ref{DOS} below). 9$\times $9$\times$9 and 5$\times $5$\times$5
Monkhorst-Pack \cite{Monk} \emph{k} point-meshes in the full wedge
of the Brillouin zone are used for ThH$_{2}$ and Th$_{4}$H$_{15}$,
respectively. Electron wave function is expanded in plane waves up
to a cutoff energy of 450 eV, and all atoms are fully relaxed until
the Hellmann-Feynman forces become less than 0.02 eV/\AA .

In present work, the theoretical equilibrium volume, bulk modulus
\emph{B}, and pressure derivative of the bulk modulus
\emph{B$^{\prime}$} are obtained by fitting the energy-volume data
in the third-order Birch-Murnaghan equation of state (EOS)
\cite{Birch}. In order to calculate elastic constants, a small
strain is applied onto the structure. For small strain $\epsilon$,
Hooke's law is valid and the crystal energy $E(V,\epsilon)$ can be
expanded as a Taylor series \cite{Nye1},
\begin{eqnarray}
E(V,\epsilon)=E(V_{0},0)+V_{0}\sum_{i=1}^{6}\sigma_{i}e_{i}+\nonumber\\
\frac{V_{0}}{2}\sum_{i,j=1}^{6}C_{ij}e_{i}e_{j}+O(\{e_{i}^{3}\}),
\end{eqnarray}
where $E(V_{0},0)$ is the energy of the unstrained system with the
equilibrium volume $V_{0}$, $\epsilon$ is strain tensor which has
matrix elements $\varepsilon_{ij}$ ($i,j$=1, 2, and 3) defined by
\begin{eqnarray}
\varepsilon_{ij}=\left(
\begin{array}
[c]{ccc}%
e_{1} & \frac{1}{2}e_{6} & \frac{1}{2}e_{5}\\
\frac{1}{2}e_{6} & e_{2} & \frac{1}{2}e_{4}\\
\frac{1}{2}e_{5} & \frac{1}{2}e_{4} & e_{3}%
\end{array}
\right)
\end{eqnarray}
and $C_{ij}$ are the elastic constants. For cubic structures, there
are three independent elastic constants ($C_{11}$, $C_{12}$, and
$C_{44}$). So, the elastic constants for fcc ThH$_{2}$ and bcc
Th$_{4}$H$_{15}$ can be calculated from three different strains
listed in the following:
\begin{eqnarray}
&\emph{$\epsilon$$^{1}$}=(\delta,\delta,\delta,0,0,0),
\emph{$\epsilon$$^{2}$}=(\delta,0,\delta,0,0,0),\nonumber\\
&\emph{$\epsilon$$^{3}$}=(0,0,0,\delta,\delta,\delta).
\end{eqnarray}
The strain amplitude $\delta$ is varied in steps of 0.006 from
$\delta$=$-$0.036 to 0.036 and the total energies $E(V,\delta)$ at
these strain steps are calculated. After obtaining elastic
constants, we can calculate bulk and shear moduli from the
Voigt-Reuss-Hill (VRH) approximations \cite{Voigt,Reuss,Hill}. The
Voigt (Reuss) bounds on the bulk modulus \emph{B$_{V}$}
(\emph{B$_{R}$}) and shear modulus \emph{G$_{V}$} (\emph{G$_{R}$})
for these two cubic crystal systems are deduced from the formulas of
elastic moduli in Ref. \cite{Hanies}. As for bct ThH$_{2}$, the six
independent elastic constants ($C_{11}$, $C_{12}$, $C_{44}$,
$C_{13}$, $C_{33}$, and $C_{66}$) can be obtained from six different
strains listed in the following:
\begin{eqnarray}
&\emph{$\epsilon$$^{1}$}=(\delta,\delta,\delta,0,0,0), \emph{$\epsilon$$^{2}$}=(\delta,0,\delta,0,0,0),\nonumber\\
&\emph{$\epsilon$$^{3}$}=(\delta,\delta,0,0,0,0),\emph{$\epsilon$$^{4}$}=(0,0,\delta,0,0,0),\nonumber\\
& \emph{$\epsilon$$^{5}$}=(0,0,0,\delta,\delta,0),
\emph{$\epsilon$$^{6}$}=(0,0,0,0,0,\delta)
\end{eqnarray}
and the formulas of elastic moduli in VRH approximations \cite{Hill}
are from Ref. \cite{Watt}. Based on Hill approximation \cite{Hill}, \emph{B}=$\frac{1}{2}(B_{R}+B_{V})$ and \emph{G}=$\frac{1}%
{2}(G_{R}+G_{V})$. The Young's modulus \emph{E} and Poisson's ratio
$\upsilon$ are given by the following formulas:
\begin{eqnarray}
E=9BG/(3B+G), \upsilon=(3B-2G)/[2(3B+G)].
\end{eqnarray}

\section{results}

\subsection{Atomic and electronic structures of thorium hydrides}
\begin{figure}[tbp]
\begin{center}
\includegraphics[width=1.0\linewidth]{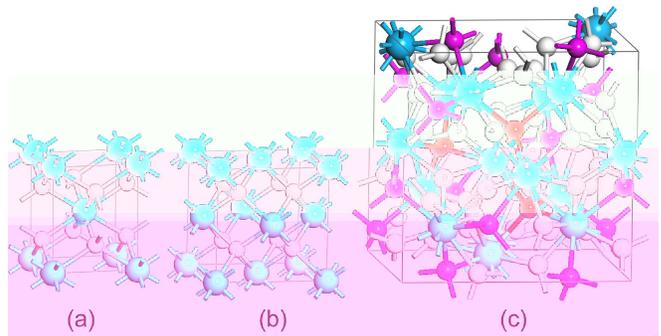}
\end{center}
\caption{(Color online) Crystal structures of (a) bct ThH$_{2}$, (b)
fcc ThH$_{2}$, and (c) bcc Th$_{4}$H$_{15}$. Here, larger cyan
spheres stand for Th atoms and the smaller white H. Note that the
magenta and red spheres in (c) stand for H1 and the white and grey
spheres H2.} \label{fig1} \label{structure}
\end{figure}%

\begin{figure}[tbp]
\begin{center}
\includegraphics[width=1.0\linewidth]{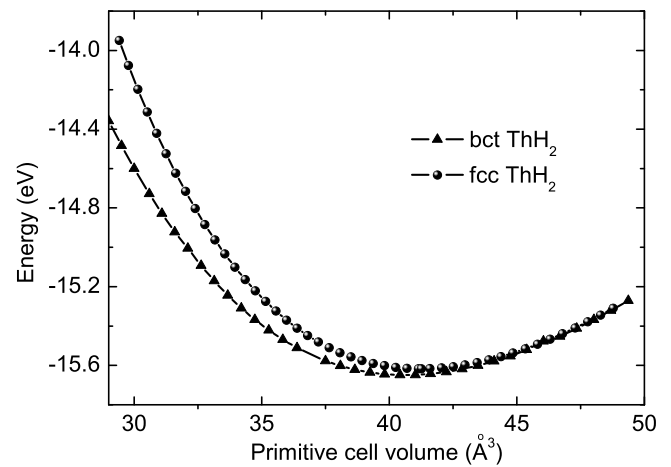}
\end{center}
\caption{Comparison of total energy vs the primitive cell volume for
ThH$_{2}$ in bct and fcc phases.} \label{energy}
\end{figure}%

\begin{table*}
\caption{Optimized structural parameters, partial and total
densities of states at the Feimi level $N$(\emph{E}$_{\rm{F}}$)
(states/eV/Th atom) for thorium hydrides.
For comparison, other theoretical results and available experimental data are also listed.}%
\begin{ruledtabular}
\begin{tabular}{lllcclll}
Compound&Cell constants ({\AA})&Coordinates&Th 6\emph{d}&Th 5\emph{f}&H \emph{s}&Total\\
\hline
bct ThH$_{2}$&\emph{a}=4.0670 (4.0475)$^{\emph{a}}$, (4.10)$^{\emph{c}}$&Th (2\emph{a}): 0, 0, 0&0.383&0.165&0.002&0.996\\
(I4/mmm)&\emph{c}=4.9125 (4.9778)$^{\emph{a}}$, (5.03)$^{\emph{c}}$&H (4\emph{d}): 0, 0.5, 0.25&(0.207)$^{\emph{a}}$&(0.140)$^{\emph{a}}$&(0.001)$^{\emph{a}}$&(0.866)$^{\emph{a}}$\\
fcc ThH$_{2}$&\emph{a}=5.4851 (5.4902)$^{\emph{a}}$,
(5.489)$^{\emph{b}}$&Th (4\emph{a}): 0, 0, 0&0.610&0.244&0.003&1.557\\
(Fm$\bar{3}$m)&&H (8\emph{c}): 0.25, 0.25, 0.25&(0.374)$^{\emph{a}}$&(0.202)$^{\emph{a}}$&(0.002)$^{\emph{a}}$&(1.451)$^{\emph{a}}$\\
Th$_{4}$H$_{15}$&\emph{a}=9.1304 (9.1280)$^{\emph{a}}$,
(9.11)$^{\emph{c,d}}$&Th (16\emph{c}): \emph{x}, \emph{x}, \emph{x};
\emph{x}=0.2087 (0.208)$^{\emph{d
}}$&0.610&0.678&0.031&2.334\\
(I$\bar{4}$3d)&&H1 (12\emph{a}): 0.375, 0, 0.25&(0.441)$^{\emph{a}}$&(0.619)$^{\emph{a}}$&(0.049)$^{\emph{a}}$&(2.474)$^{\emph{a}}$\\
&&H2 (48\emph{e}): 0.372, 0.219, 0.404 (0.4, 0.23, 0.372)$^{\emph{d
}}$&\\
\end{tabular}
\end{ruledtabular}
\label{structure} $^{\emph{a}}$ Reference \cite{Shein},
$^{\emph{b}}$ Reference \cite{Brooks}, $^{\emph{c}}$ Reference
\cite{Keller}, $^{\emph{d}}$ Reference \cite{Zachariasen}.
\end{table*}

\begin{figure*}[tbp]
\begin{center}
\includegraphics[width=1.0\linewidth]{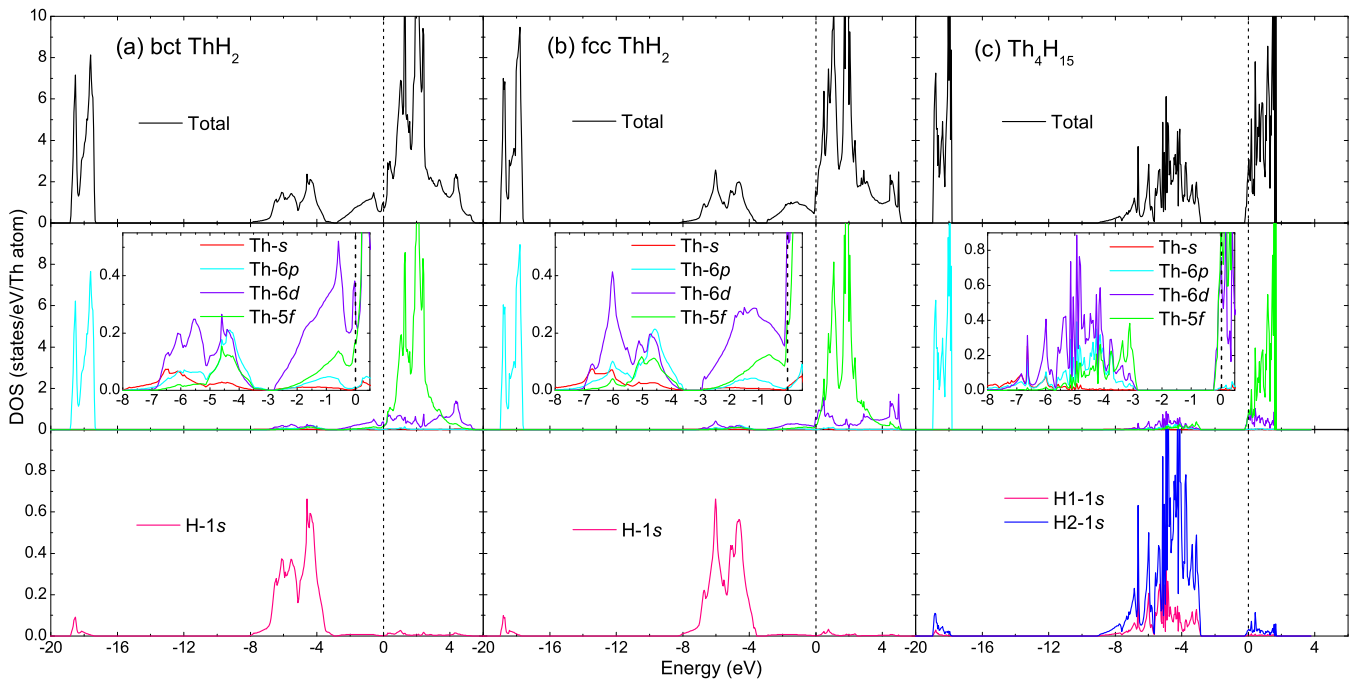}
\end{center}
\caption{(Color online) Total and orbital-resolved local densities
of states for (a) bct ThH$_{2}$, (b) fcc ThH$_{2}$, and (c) bcc
Th$_{4}$H$_{15}$. The Fermi energy level is set at zero.}
\label{DOS}
\end{figure*}%
At ambient condition, the stable thorium dihydride crystallizes in a
body-centered tetragonal (bct) ionic structure with space group
\emph{I\rm{4}}/\emph{mmm} (No. 139). Its unit cell is composed of
two ThH$_{2}$ formula units with the thorium atoms and the hydrogen
atoms in 2\emph{a} (in Wyckoff notation) and 4\emph{d} sites,
respectively [see Fig. 1(a)]. Each Th atom is surrounded by eight H
atoms forming a tetragonal and each H connects with four Th atoms to
build a tetrahedron. The present optimized lattice parameters
(\emph{a} and \emph{c}) are 4.0670 \AA\ and 4.9125 \AA\ (see Table
\ref{structure}), in good agreement with the experimental
\cite{Keller} values of 4.10 \AA\ and 5.03 \AA. Besides, the
face-centered cubic (fcc) fluorite-type structure with space group
\emph{Fm$\bar{3}$m} (No. 225) is considered as metastable phase for
ThH$_{2}$ [see Fig. 1(b)]. The fluorite structure is the stable
structure of all actinide dioxides. Moreover, this fcc structure can
be obtained from the bct stable structure by modulating the base
vectors. In fcc structure, each Th atom is surrounded by eight H
atoms forming a cube and each H connects with four Th atoms to build
a tetrahedron. Our optimized lattice constant (\emph{a}) for fcc
ThH$_{2}$ is 5.4851 \AA\ (see Table \ref{structure}), in excellent
agreement with the measured \cite{Brooks} values of 5.489 \AA. The
structure of Th$_{4}$H$_{15}$ is more complicated. It crystallizes
in a body-centered cubic (bcc) structure with space group
\emph{I$\bar{4}3$d} (No. 220) [see Fig. 1(c)]. The unit cell has
sixteen Th (16\emph{c}), twelve H1 (12\emph{a}), and forty eight H2
(48\emph{e}) atoms. Each H1 is surrounded by four Th atoms to build
the tetrahedral structure and each H2 is located at the center of
triangle formed by three Th atoms. As a result, each Th is
surrounded by three H1 atoms and nine H2 atoms. In Fig. 1(c), we
label clearly the H1 atoms (red spheres) and H2 atoms (grey spheres)
connected to the Th atom located at center. The current optimized
structural parameters of Th$_{4}$H$_{15}$ are listed in Table
\ref{structure}, where one can find that our calculation results are
in good agreement with experiment \cite{Zachariasen}.

To investigate the stability of ThH$_{2}$, we have optimized the
structural parameters of its bct and fcc structures at different
pressures. To avoid the Pulay stress problem, we perform the
structure relaxation calculations at fixed volumes rather than
constant pressures. For fcc structure, due to its high symmetry, the
structure relaxation calculations are performed at fixed volumes
with no relaxation of coordinates and cell shape. However, for bct
structure, the cell shape is necessary to be optimized due to their
internal degrees of freedom. The total energies (per primitive cell)
of the two structures at different volumes are calculated and shown
in Fig. \ref{energy}. Obviously, the bct ThH$_{2}$ is more stable
than fcc ThH$_{2}$ under ambient pressure. The equilibrium volumes
of bct and fcc structures are 40.63 {\AA}$^{3}$ and 41.26
{\AA}$^{3}$, respectively. Thus the equilibrium volumes of both bct
and fcc phases are approximately equal and the total energy
difference at their equilibrium states is $\mathtt{\sim}$0.032 eV.
This conclusion is consistent well with the previous FLAPW results
\cite{Shein}.

Basically, all the macroscopical properties of materials, such as
hardness, elasticity, and conductivity, originate from their
electronic structure properties as well as chemical bonding nature.
Therefore, it is necessary to perform the electronic structure
analysis of thorium hydrides. The calculated total densities of
states (DOS) and the orbital-resolved partial densities of states
(PDOS) of bct ThH$_{2}$, fcc ThH$_{2}$ and bcc Th$_{4}$H$_{15}$ are
shown in Fig. \ref{DOS}. Overall, the occupation properties of
ThH$_{2}$ in bct and fcc phases are similar. The occupied DOS near
the Fermi level is featured by the three well-resolved peaks. The
one near $-$1.0 eV is principally Th $6d$ in character, while the
other two peaks respectively near $-$4.5 and $-$6.0 eV are mostly H
$1s$ hybridized with Th $6d$ and Th $5f$ orbitals. These
well-separated orbital peaks have been observed in the photoelectron
spectroscopy (PES) measurement \cite{Weaver}. In addition, the band
width of the Th $6d$ valence band near the Fermi level is 3 eV,
consistent with the FLAPW calculation \cite{Shein} and experimental
data \cite{Weaver}. The H $1s$ valence band width is of 5.0 eV, also
in accord with the experimental data (6.0 eV) and previous FLAPW
result (5.2 eV). Moreover, the low bands of bct (fcc) ThH$_{2}$
covering from $-$18.8 ($-$19.0) to $-$17.2 ($-$17.5) eV is mainly
featured by Th $6p$ state mixed with a little H $1s$ state and the
conduction bands is principally occupied by Th $5f$ states with
admixtures of the Th 6\emph{d} and H 1\emph{s} states and has a
width of 5.5 (5.1) eV. For Th$_{4}$H$_{15}$, our calculations
reproduce all the features calculated by Shein \emph{et al}.
\cite{Shein} and agree well with PES measurements \cite{Weaver}.
Note that our calculated conduction band width is 1.7 eV, consistent
with the PES experimental value \cite{Weaver} (0.9 eV) and evidently
smaller than that of FLAPW calculation result \cite{Shein}. The main
occupation of conduction bands is Th $5f$ orbital, mixed with a
little Th $6d$ and H $1s$ states. In addition, we have also
presented in Table \ref{structure} the partial and total DOSs at the
Feimi level $N$(\emph{E}$_{\rm{F}}$) for thorium hydrides to study
the occupation of the conduction band. Clearly, our results are in
good agreement with the FLAPW calculation \cite{Shein}.

\begin{figure}[tbp]
\begin{center}
\includegraphics[width=1.0\linewidth]{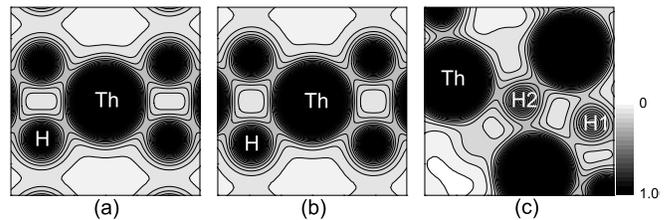}
\end{center}
\caption{Valence charge density of (a) bct ThH$_{2}$ in (100) plane,
(b) fcc ThH$_{2}$ in (1$\bar{1}$0) plane, and (c) bcc
Th$_{4}$H$_{15}$ in plane established by three Th and one H2 atoms.
Contour lines are drawn from 0.0 to 1.0 at 0.05 e/{\AA}$^{3}$
intervals.} \label{charge}
\end{figure}

\begin{figure}[tbp]
\begin{center}
\includegraphics[width=1.0\linewidth]{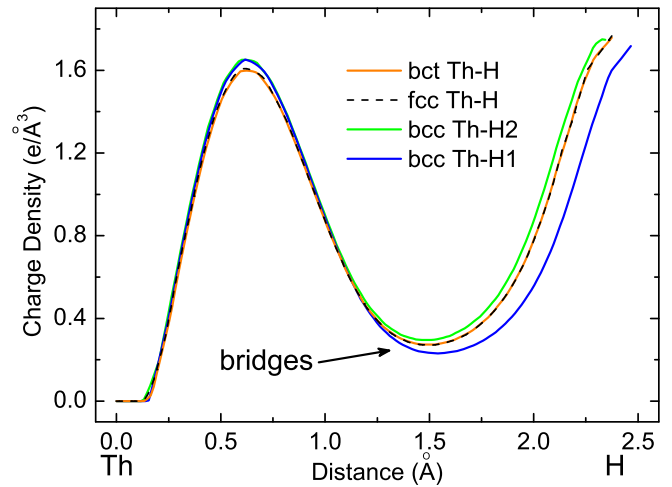}
\end{center}
\caption{(Color online) Line charge density distribution between Th
atom and the nearest neighbor H atom for bct ThH$_{2}$, fcc
ThH$_{2}$, and bcc Th$_{4}$H$_{15}$.} \label{linecharge}
\end{figure}%

\begin{table*}
\caption{Calculated charges and volumes according to Bader
partitioning as well as the Th$-$H distances and correlated minimum
values of charge densities along the Th$-$H bonds for thorium
hydrides. As a comparison, other
theoretical work and available experimental data are also listed. Note that some properties of Th$_{4}$H$_{15}$ have two values (first for H1 and second for H2).}%
\begin{ruledtabular}
\begin{tabular}{cccccccccccccc}
Compound&Q$_{B}$(Th)&Q$_{B}$(H)&V$_{B}$(Th)&V$_{B}$(H)&Th$-$H&Charge density$_{\rm{min.}}$\\
&($e$)&($e$)&({\AA}$^{3}$)&({\AA}$^{3}$)&({\AA})&($e$/{\AA}$^{3}$)\\
\hline
bct ThH$_{2}$&10.530&1.735&25.59&7.50&2.375 (2.385)$^{\emph{a}}$&0.273\\
fcc ThH$_{2}$&10.488&1.756&25.72&7.77&2.375 (2.337)$^{\emph{b}}$&0.271\\
Th$_{4}$H$_{15}$&9.870&1.590, 1.563&20.15&7.16, 7.35&2.465 (2.46)$^{\emph{a}}$, 2.343 (2.29)$^{\emph{a}}$&0.231, 0.296\\
\end{tabular}
\end{ruledtabular}
\label{bader} $^{\emph{a}}$ Reference \cite{Keller}, $^{\emph{b}}$
Reference \cite{Brooks}.
\end{table*}

In order to gain more insight into the bonding nature of ground
state thorium hydrides, we also investigate the valence charge
density distribution. The calculated valence charge density maps of
the bct ThH$_{2}$ in (100) plane, fcc ThH$_{2}$ in (1$\bar{1}$0)
plane, and bcc Th$_{4}$H$_{15}$ in plane established by three Th and
one H2 atoms are plotted in Fig. \ref{charge}. Clearly, the charge
densities around Th and H ions are all near spherical distribution
with slightly deformed towards the direction to their nearest
neighboring atoms. To describe the ionic/covalent character
quantitatively and more clearly, we plot in Fig. \ref{linecharge}
the line charge density distribution along the nearest Th$-$H bonds
and perform the Bader analysis \cite{Bader,Tang}. For bct and fcc
ThH$_{2}$, the two charge density curves change with the same way
along the Th$-$H bonds. For Th$_{4}$H$_{15}$, the two curves along
the Th$-$H1 and Th$-$H2 bonds are almost the same before the bridge
locus (indicated by the arrow), after where they split evidently.
Besides, one can find a minimum value of charge density in Fig.
\ref{linecharge} for each bond at around the bridge. These minimum
values are listed in Table \ref{bader}. Although these values are
much smaller than 0.7 $e$/{\AA}$^{3}$ for Si covalent bond, they are
prominently higher than 0.05 $e$/{\AA}$^{3}$ for Na$-$Cl bond in
typical ionic crystal NaCl. Also we notice that these values are
smaller than the minimum values (0.45 e/{\AA}$^{3}$) of charge
density along the Th$-$O bond in our previous study of ThO$_{2}$
\cite{Wang}. Therefore, there are weak but clear covalent bonds
between Th and H for ThH$_{2}$ in bct and fcc phases and for bcc
Th$_{4}$H$_{15}$. The slight difference appears in Th$_{4}$H$_{15}$,
where the minimum value of charge density in Th$-$H1 bond is smaller
than that in Th$-$H2 bond. In addition, one can notice from Table
\ref{bader} that our calculated Th$-$H distances are all well
consistent with the previous theoretical \cite{Brooks} and
experimental \cite{Keller} results. As for Bader analysis, it is a
well established analysis tool for studying the topology of the
electron density and thus is particularly suitable for discussing
the mixed ionic/covalent character of a compound. The charge
(Q$_{B}$) enclosed within the Bader volume (V$_{B}$) is a good
approximation to the total electronic charge of an atom. In the
present study, the default charge density grids for one unit cell
are 56$\times$56$\times$72, 56$\times$56$\times$56, and
112$\times$112$\times$112 for bct ThH$_{2}$, fcc ThH$_{2}$, and bcc
Th$_{4}$H$_{15}$, respectively. To check the precision, the charge
density distribution for fcc ThH$_{2}$ is calculated with a series
of \emph{n} times finer grids (\emph{n} = 2, 3, 4, 5, 6). The
deviation of the effective charge between the five and the six times
finer grids is less than 0.02\%. Thus we perform the charge density
calculations using the six times finer grid
(336$\times$336$\times$432, 336$\times$336$\times$336, and
672$\times$672$\times$672 for bct ThH$_{2}$, fcc ThH$_{2}$, and bcc
Th$_{4}$H$_{15}$, respectively). The calculated results are
presented in Table \ref{bader}. Note that although we have included
the core charge in charge density calculations, since we do not
expect variations as far as the trends are concerned, only the
valence charge are listed. From Table \ref{bader} the following
prominent features can be seen: (i) The Bader charges and volumes
for bct and fcc ThH$_{2}$ are almost equal to each other. This shows
the same ionic character, through a flux of charge (about 1.5
electrons for each Th atom) from cations towards anions, for thorium
dihydride both in their stable phase and metastable phase.  (ii) The
Bader charges and volumes for bcc Th$_{4}$H$_{15}$ are different
comparing to thorium dihydride. This originates from their different
hydrogen concentration. For bcc Th$_{4}$H$_{15}$, about 2.1
electrons transfer from each Th atom to H. This indicates that the
Th atoms in Th$_{4}$H$_{15}$ are more ionized than that in thorium
dihydrides.

\begin{table*}
\caption{ Calculated elastic constants, various moduli, and
Poisson's ratio $\upsilon$ for thorium hydrides.}%
\begin{ruledtabular}
\begin{tabular}{cccccccccccccc}
Compound&\emph{C$_{11}$}&\emph{C$_{12}$}&\emph{C$_{44}$}&\emph{C$_{13}$}&\emph{C$_{33}$}&\emph{C$_{66}$}&\emph{B}&\emph{B$^{'}$}&\emph{G}&\emph{E}&$\upsilon$\\
&(GPa)&(GPa)&(GPa)&(GPa)&(GPa)&(GPa)&(GPa)&(GPa)&(GPa)&(GPa)&\\
\hline
bct ThH$_{2}$&131&100&29&78&89&5&91&3.3&16&44&0.419\\
fcc ThH$_{2}$&46&119&57&&&&95 (99)$^{\emph{a}}$&3.8&&&\\
Th$_{4}$H$_{15}$&121&66&59&&&&85&2.2&43&111&0.282\\
\end{tabular}
\end{ruledtabular}
\label{elastic} $^{\emph{a}}$ Reference \cite{Brooks}.
\end{table*}

\subsection{Mechanical properties of thorium hydrides}%
Our calculated elastic constants, various moduli, pressure
derivative of the bulk modulus \emph{B$^{^{\prime}}$}, and Poisson's
ratio $\upsilon$ for bct ThH$_{2}$, fcc ThH$_{2}$, and bcc
Th$_{4}$H$_{15}$ are collected in Table \ref{elastic}. Obviously,
bct ThH$_{2}$ is mechanically stable due to the fact that its
elastic constants satisfy the following mechanical stability
criteria \cite{Nye2} of tetragonal structure:
\begin{eqnarray}
&  C_{11}>0, C_{33}>0, C_{44}>0, C_{66}>0,\nonumber\\
&  (C_{11}-C_{12})>0, (C_{11}+C_{33}-2C_{13})>0,\nonumber\\
&  [2(C_{11}+C_{12})+C_{33}+4C_{13}]>0.
\end{eqnarray}
Th$_{4}$H$_{15}$ is also mechanically stable because its elastic
constants satisfy the following mechanical stability criteria
\cite{Nye2} of cubic structure:
\begin{eqnarray}
C_{11}>0, C_{44}>0, C_{11}>|C_{12}|, (C_{11}+2C_{12})>0.
\end{eqnarray}
However, the ThH$_{2}$ in its metastable cubic phase is mechanically
unstable. In fact, one can see from Table \ref{elastic} that
$C_{11}$ is much smaller than $C_{12}$ for fcc ThH$_{2}$. Therefore,
the mechanical stability criteria \cite{Nye2}
$C_{11}$$>$$\mid$$C_{12}$$\mid$ can not be satisfied and the shear
modulus, Young's modulus, and Poisson's ratio also can not be
obtained. As for bulk modulus \emph{B}, each derived value from VRH
approximations \cite{Hill} for all the thorium hydrides turns out to
be very close to that obtained by fitting Murnaghan EOS
\cite{Birch}. Moreover, we notice that our calculated \emph{B} for
fcc ThH$_{2}$ is approximately equal to previous LMTO theoretical
result \cite{Brooks}. For shear modulus \emph{G} and Young's modulus
\emph{E}, their values for bcc Th$_{4}$H$_{15}$ are approximately
2.5 times of those for bct ThH$_{2}$. For Poisson's ratio, the
values for both bct ThH$_{2}$ and bcc Th$_{4}$H$_{15}$ are within
the range from 0.25 to 0.45 for typical metals.

\subsection{Phonon dispersion curves of thorium dihydrides}%
\begin{figure*}[ptb]
\begin{center}
\includegraphics[width=0.8\linewidth]{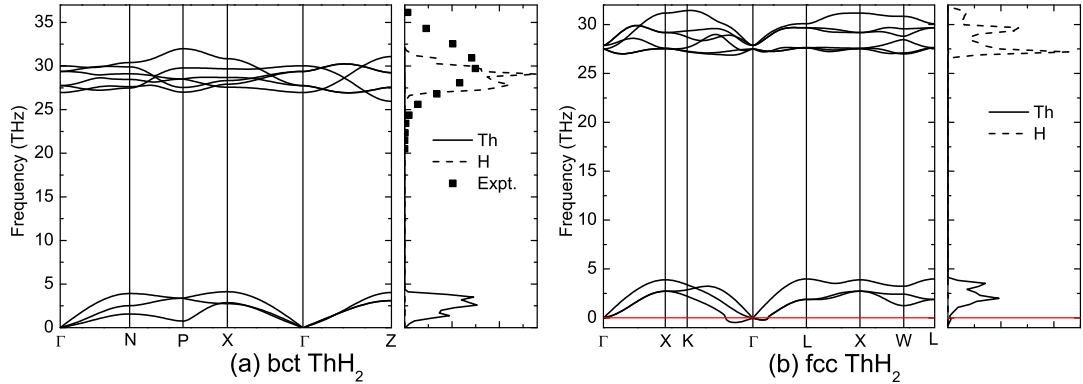}
\end{center}
\caption{(Color online) Calculated phonon dispersion curves and
corresponding PDOS
for (a) bct ThH$_{2}$ and (b) fcc ThH$_{2}$. For comparison, the experimental data in Ref. [4] are also shown (squares).}%
\label{phonon}\label{phonon}
\end{figure*}

Employing the Hellmann-Feynman theorem and the direct method, we
have calculated the phonon curves along some high-symmetry
directions in the Brillouin zone (BZ), together with the phonon
PDOS. For the phonon dispersion calculation, we use the
2$\times$2$\times$2 bct (fcc) supercell containing 48 (96) atoms for
bct (fcc) ThH$_{2}$. To calculate the Hellmann-Feynman forces, we
displace four and eight atoms, respectively, for bct and fcc
ThH$_{2}$ from their equilibrium positions and the amplitude of all
the displacements is 0.03 \AA. The calculated phonon dispersion
curves along the $\Gamma$$-$$N$$-$$P$$-$$X$$-$$\Gamma$$-$$Z$
directions for bct ThH$_{2}$ and along the
$\Gamma$$-$$X$$-$$K$$-$$\Gamma
$$-$$L$$-$$X$$-$$W$$-$$L$ directions for fcc ThH$_{2}$ are displayed in Fig. \ref{phonon}(a) and Fig. \ref{phonon}(b), respectively.

For both bct and fcc ThH$_{2}$, there are only three atoms in their
primitive cells. Therefore, nine phonon modes exist in the
dispersion relations. Due to the fact that thorium is much heavier
than hydrogen atom, the vibration frequency of thorium atom is
apparently lower than that of hydrogen atom. Therefore, evident gap
between the optic modes and the acoustic branches exits and the
phonon DOS of bct (fcc) ThH$_{2}$ can be viewed as two parts. One is
the part lower than 4.4 (4.1) THz where the main contribution comes
from the thorium sublattice, while the other part range from 26.0 to
32.1 (26.4 to 31.8) THz is dominated by the dynamics of the light
hydrogen atoms. In experimental measurements, Dietrich \emph{et al.}
reported that the acoustic branches of ThH$_{2}$ are lower than 4.8
THz and optic modes range from 23.4 to 36.2 THz \cite{Dietrich}.
Their measured data of optic modes are presented in Fig.
\ref{phonon}(a) for comparison and the acoustic-phonon data are
unable obtained from their time-of-flight spectrum. Obviously, our
calculated results are on the whole consistent with the experimental
data. In addition, while all frequencies are positive for bct
ThH$_{2}$, which assures a dynamical stability of the bct phase
against mechanical perturbations. For fcc ThH$_{2}$, one can see
from Fig. \ref{phonon} (b) that the transverse acoustic (TA) mode
close to $\Gamma$ point becomes imaginary along the $\Gamma$$-$$K$
and $\Gamma$$-$$L$ directions. The minimum of the TA branch occurs
along the $\Gamma$$-$$K$ direction. This indicates instability of
fcc phase of ThH$_{2}$ compared to stable bct phase, which is well
consistent with our previous mechanical stability analysis of
ThH$_{2}$. Moreover, there exists LO-TO splitting at $\Gamma$ point
in bct phase. But in fcc phase there is no LO-TO splitting.

\section{CONCLUSION}
In summary, we have used the first-principles DFT-GGA method to
calculate the structural, electronic, mechanical, and thermodynamic
properties of ThH$_{2}$ in its stable bct phase and metastable fcc
phase and Th$_{4}$H$_{15}$ in bcc phase. Our optimized structural
parameters are well consistent with experiments. The occupation
characters of electronic orbitals also accord well with experiments
and previous calculations. Through Bader analysis, we have found
that the Th$-$H bonds in all thorium hydrides exhibit weak covalent
character, but the ionic property for ThH$_{2}$ and Th$_{4}$H$_{15}$
are different. While $\mathtt{\sim}$1.5 electrons transfer from each
Th atom to H in ThH$_{2}$, in Th$_{4}$H$_{15}$ about 2.1 electrons
deviate from each Th atom. In addition, our calculated phonon curves
of fcc ThH$_{2}$ have shown that the TA mode becomes imaginary close
to $\Gamma$ point and the subsequent instability of this phase.

\begin{acknowledgments}
This work was supported by the Foundations for Development of
Science and Technology of China Academy of Engineering Physics under
Grants No. 2009B0301037 and No. 2009A0102005.
\end{acknowledgments}


\begin{thebibliography}{99}                                                                                               %

\bibitem {Satterthwaite1}C. B. Satterthwaite and I. L. Toepke,
Phys. Rev. Lett. \textbf{25}, 741 (1970).

\bibitem {Shein}I. R. Shein, K. I. Shein, N. I. Medvedeva, and A.
L. Ivanovskii, Physica B \textbf{389}, 296 (2007).

\bibitem {Satterthwaite2}C. B. Satterthwaite and D. T. Peterson, J.
Less-Common Met. \textbf{26}, 361 (1972).

\bibitem {Dietrich}M. Dietrich, W. Reichardt, and H. Rietschel,
Solid State Commun. \textbf{21} 603 (1977).

\bibitem {Brooks}M. S. S. Brooks and B. Johansson, Physica B
\textbf{130}, 516 (1985).

\bibitem {Bader}R. F. W. Bader, \emph{Atoms in Molecules: A Quantum
Theory} (Oxford University Press, New York, 1990).

\bibitem {Tang}W. Tang, E. Sanville, and G. Henkelman, J. Phys.:
Condens. Matter \textbf{21}, 084204 (2009).

\bibitem {Kresse3}G. Kresse and J. Furthm\"{u}ller, Phys. Rev. B
\textbf{54}, 11169 (1996).


\bibitem {GGA}J. P. Perdew, K. Burke, and Y. Wang, Phys. Rev. B
\textbf{54}, 16533 (1996).

\bibitem {PAW}P. E. Bl\"{o}chl, Phys. Rev. B \textbf{50}, 17953
(1994).

\bibitem {Monk}H. J. Monkhorst and J. D. Pack, Phys. Rev. B
\textbf{13}, 5188 (1972).

\bibitem {Birch}F. Birch, Phys. Rev. \textbf{71}, 809 (1947).

\bibitem {Nye1}J.F. Nye, Physical Properties of Crystals,
Their Representation by Tensors and Matrices, Oxford Press, Chap.
VIII, 1957.

\bibitem {Voigt}W. Voigt, \emph{Lehrburch der Kristallphysik}
(Teubner, Leipzig, 1928).

\bibitem {Reuss}A. Reuss and Z. Angew, Math. Mech. \textbf{9}, 49
(1929).

\bibitem {Hill}R. Hill, Phys. Phys. Soc. London \textbf{65}, 349 (1952).

\bibitem {Hanies}J. Hanies, J. M. Leger, and G. Bocquillon, Annu. Rev. Mater. Res. \textbf{31}, 1 (2001).

\bibitem {Watt}J. P. Watt and L. Peselnick, J. Appl. Phys.
\textbf{51}, 1525 (1980).

\bibitem {Keller}C. Keller, \textit{Thorium} (Springer, 1978).

\bibitem {Zachariasen}W. H. Zachariasen, Acta Cryst. \textbf{6},
395 (1953).

\bibitem {Weaver}J. H. Weaver, J. A. Knapp, D. E. Eastman, D. T.
Peterson, and C. B. Satterthwaite, Phys. Rev. Lett. \textbf{39}, 639
(1977).

\bibitem {Wang}B. Wang, H, Shi, W. Li, and P. Zhang,
aiXiv:0908.3558v1 [cond-mat.mtrl-sci] (2009).

\bibitem {Nye2}J. F. Nye, \emph{Physical Properties of Crystals}
(Oxford University Press, Oxford, 1985).



\end{thebibliography}
\end{document}